\documentclass[a4paper]{ESASPCS13Style}
\usepackage{epsfig}
\include{latex_macros}

\begin{document}
\newcommand{\vsini}{\mbox{$v_e\,\sin\,i$}}
\newcommand{\kmsec}{\,\mbox{$\mbox{km}\,\mbox{s}^{-1}$}}
\newcommand{\rstar}{\,\mbox{$\mbox{R}_*$}}

\title{Inferring coronal structure using X-ray spectra: a {\em Chandra} study of AB Dor}

\author{G.A.J. Hussain\inst{1,2}, N.S. Brickhouse\inst{2}, A.K. Dupree\inst{2},
M. Jardine\inst{3}, A. van Ballegooijen\inst{2}, A. Collier Cameron\inst{3}, 
J.-F. Donati\inst{4} \and  F. Favata\inst{1} 
  \institute{Astrophysics Division -- Research and
  Space Science Department of ESA, ESTEC, Postbus 299, NL-2200 AG
  Noordwijk, The Netherlands \and 
  Harvard Smithsonian CfA -- Cambridge MA 02138, USA \and 
  University of St Andrews -- St Andrews, KY16 9SS, Scotland 
  \and Observatoire Midi-Pyrenees -- Meudon, France }}

\maketitle 
\begin{abstract}
The {\em Chandra} X-ray observatory monitored the single cool star,
AB Doradus, continuously for a period lasting 88\,ksec
(1.98\,$P_{\rm rot}$)  in 2002 December with the LETG/HRC-S.
The X-ray lightcurve shows significant rotational modulation.
It can be represented as having a flat level of emission superimposed
with bright flaring regions that appear at the same phases in both
rotation cycles. Phase-binned O\,{\sc viii} line profiles show centroid shifts
that also repeat in consecutive rotation cycles. These  
Doppler shifts trace a roughly sinusoidal pattern with a 
a semi-amplitude of $30 \pm 10$\kmsec.
By taking both the lightcurve and spectral diagnostics into account 
along with constraints   on the rotational broadening of line profiles
(provided by archival  Chandra HETG  {Fe}\,{\sc xvii}
line profiles) we can construct a simple model of the X-ray
corona. The corona can be described as having two components, 
one component is homogeneously distributed, extending less than 1.75\rstar; 
and the other consists of at least two compact emitting
regions near the stellar surface. 
These  compact regions account for 16\% of the
X-ray emission and are likely to be 
located less than 0.4\rstar\ above the stellar surface.

\keywords{Stars: structure  -- Stars: coronae -- Stars: X-rays -- Stars: spectroscopy }
\end{abstract}

\section{Introduction}
 
While the thermal properties of active stellar coronae are increasingly well-determined,
we have yet to establish where the emitting plasma is located. As active stars
cannot be spatially resolved at X-ray wavelengths we have to rely on indirect
techniques to infer the distribution of coronal material. 
Grating spectra from the {\em Chandra}/LETG configuration 
are sufficiently resolved to measure radial velocities in stellar coronae (Brickhouse et al. 2001; 
Hoogerwerf et al. 2004).

Previous studies of AB Dor data taken using ROSAT show evidence of
rotational modulation at the 5--13\% level (K\"urster et al. 1997).
We set out to study coronal structure in the active single K0 dwarf, AB Dor 
($P_{rot}=0.51$\,d, $v_e \sin i = 90$\,km\,s$^{-1}$) by analysing {\em Chandra}/LETG
data covering almost two rotation cycles.
We find strong evidence for rotational modulation in both the spectra and lightcurves and use
the results to develop a simple model for the star's X-ray emitting corona.
Section~2 summarises the properties of the LETG X-ray lightcurve, Section~3
details measurements of rotational modulation in the LETG spectra and Section~4 
shows the upper limits to which the bulk of the X-ray emitting corona extends, as
determined using archive {\em Chandra}/HETG spectra. The main conclusions and 
a simple coronal model are presented in Section~5.

\section{Lightcurves: detecting rotational modulation}

\begin{figure*}[bht]
  \begin{center}
    \epsfig{file=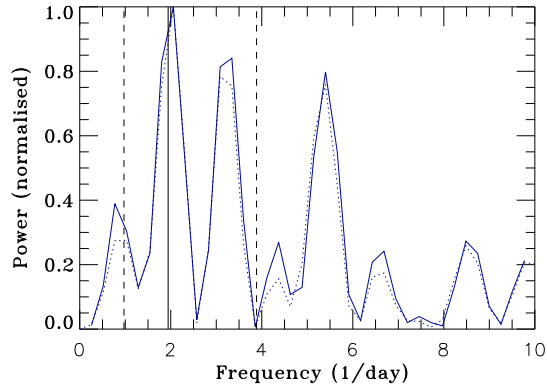, width=8cm}

  \end{center}
\caption{Period analysis of the LETG lightcurve. 
The solid line represents the Lomb-Scargle periodogram and the dotted line
shows the CLEAN-like analysis. Clearly the strongest peak corresponds 
with the stellar rotation period. Given the width of the strongest power peak
this value is absolutely consistent with AB Dor's period.
\label{fig:light1}}
\end{figure*}

\begin{figure*}[bht]
  \begin{center}
    \epsfig{file=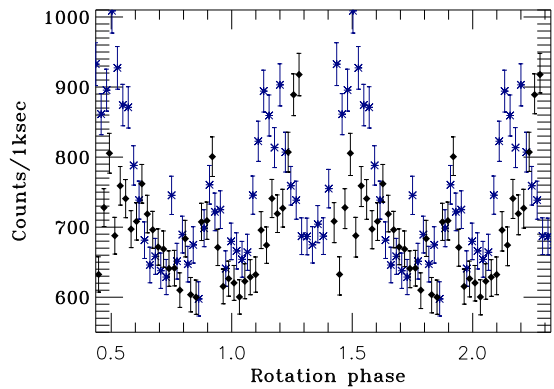, width=8cm}
  \end{center}
\caption{ The LETG lightcurve folded on AB Dor's rotation period (0.51\,d).
The ephemeris used is from Innis et al. (1988).
Blue asterisks show the first cycle and diamonds  represent the second rotation cycle.
Two complete cycles are not obtained, causing the gap in phase
coverage between phases 0.3 and 0.4 in the second cycle.\label{fig:light2}}
\end{figure*}

The  lightcurve is
extracted over the short wavelength region ($1 < \lambda < 50$\,\AA) of the dataset,
using 1\,ksec bins, summing up both the $+1$ and $-1$ orders in order to
ensure effective subtraction of the X-ray background.
Periodogram analyses show that there is strong evidence for
rotational modulation, with the strongest peak occurring close
to AB Dor's rotation period (Figure\,\ref{fig:light1}). Given the width of the peak, 
we find that this is consistent with AB Dor's period.
By folding the lightcurve on AB Dor's rotation period (Figure\,\ref{fig:light2})
it is evident that there is a high level of agreement between the two
consecutive rotation cycles. By fitting sine-curves to the lightcurves, we measure 
rotational modulation at the 12\% level. 

\section{Spectra: measuring centroid shifts}
\begin{figure*}[ht]
  \begin{center}
    \epsfig{file=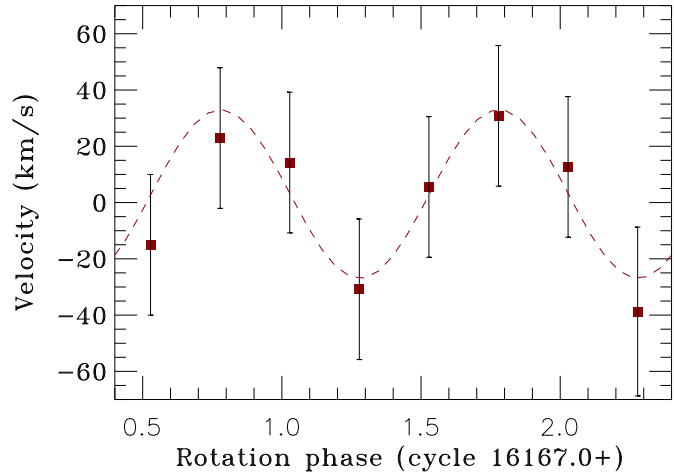, width=10cm}
  \end{center}
\caption{Velocity shifts in the line centroids of the 
O\,{\sc viii}\,18.97\,\AA\ profile (red squares), a sine
wave is overplotted (red dotted line). Two consecutive
phases were observed.
\label{fig:spectra}}
\end{figure*}

The absolute wavelength scale is not important as we are 
looking for {\em relative shifts} in the positions of the line
centroids as a function of rotation phase. To ensure accurate calibration of the wavelength
positions of the strongest lines,  Gaussians are fitted to to the line profiles
(integrated over the entire 88.1\,ksec exposure).
Offsets between the centroid positions of the line 
profiles and the laboratory wavelengths are computed
and used to recalibrate their ``zero-velocity'' positions.
The total exposure is divided into eight quarter rotation-phase bins
and spectra are extracted for each of these bins. 
Line centroids in these eight sets of spectra are re-fitted using Gaussians
in order to evaluate any Doppler shifts. 
Note that all
the bins are of 11.12\,ksec length except for the last phase bin which is
slightly shorter, 10.26\,ksec (as the total observation
does not cover two full rotation cycles).
Because the dither time-scale of the spacecraft is 1087\,sec,
 each phase bin consists of approximately ten dither cycles.
Thus the centroid measurements should not be susceptible to wavelength
deviations associated with dithering.
We find that the O{\sc viii} 18.97\AA\ line profile can be fitted with greatest precision
due to its relative strength and lack of blends.
Any offsets are then measured relative to the zero-velocity
positions (measured using the full 88.1\,ksec exposure), 
and mean Doppler shifts are obtained by combining  the $+1$ and $-1$ orders.
We find velocity shifts that repeat from in two consecutive cycles
(Figure\,\ref{fig:spectra}).

\section{Spectra: Evaluating rotational broadening}

\begin{figure*}[ht]
  \begin{center}
    \epsfig{file=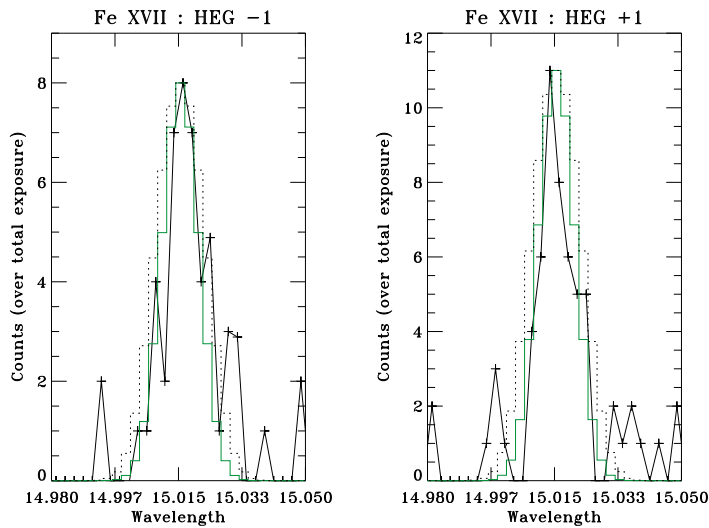,width=10cm}
  \end{center}
\caption{ Evaluating rotational broadening using archive
{\em Chandra}/HETG spectra. The solid green line shows the observed
line profiles for the $+1$ and $-1$ orders.
The green and dotted line profiles represent 
instrumental broadening convolved with 
and rotational broadening corresponding to 
 $v_e \sin i=90$\,km\,s$^{-1}$ and $180$\,km\,s$^{-1}$ 
 respectively.
\label{fig:spectrarot}}
\end{figure*}

AB Dor's photospheric lines are substantially
rotationally broadened (\vsini$=90$\,\kmsec).     
If the X-ray emitting corona is a diffuse shell extended out to several \rstar,  LETG  
line profiles at long wavelengths
should be significantly broader than profiles subject to
instrumental and thermal broadening effects alone.
Unfortunately, the LETG line profiles in the long wavelength, high resolution  
region of the dataset cannot be used to accurately measure rotational broadening, 
as the line profiles from the $+1$ and $-1$ orders are inconsistent 
(also see Chung et al. 2004). We therefore use
HETG spectra of AB Dor to evaluate rotational broadening in 
X-ray line profiles (and thus the extent of the emitting corona).
The Fe {\sc xvii} line is analysed as it is subject to less thermal
broadening than the O\,{\sc viii} line. Broadening with a 
$v_e \sin i$ of 180\,km\,s$^{-1}$ (i.e. corresponding to a corona
that extends to $\sim$2\,R$_*$) produces a profile that is broader
than the observed line (Figure\,\ref{fig:spectrarot}). 
Further analysis indicates that the bulk of the emitting corona
does not extend beyond 1.75\,R$_*$ (see Hussain et al. 2004 for more detail).

The solid black lines represent the observed HETG spectra 
in the $+1$ and $-1$ orders respectively. 
The solid and dotted histogram line profiles 
show the effects of instrumental, thermal and rotational broadening
with $v_e \sin i$ values of 90 and 180\,km\,s$^{-1}$ respectively.

\section{Conclusions}
Our main conclusions can be itemised as follows.
\begin{itemize}

\item The X-ray lightcurve has a flat level of emission superimposed
with two (or three) bright/flaring regions; one region lies between phases, $0.8<\phi<1.0$ and
one or two regions lie between $0.1<\phi<0.7$. These bright flaring regions appear to  
 repeat in both (consecutive) rotation cycles observed.

\item Phase-binned O\,{\sc viii} line profiles show centroid shifts that
also repeat from cycle to cycle and are likely to be caused by one
or more compact emitting regions. If only one region is reponsible for
this modulation, it must be caused by a compact mid-to-high latitude region.

\item We measure  a strong upper limit for rotational broadening using
archive {\em Chandra}/HETG spectra. Taking instrumental and thermal
broadening into account, this corresponds to a  
coronal $v_e \sin i < 180$\,km\,s$^{-1}$ (i.e. $R_{cor}<2$\rstar).

\end{itemize}
A simple model of AB Dor's X-ray emitting corona consistent with the above
diagnostics would have a homogeneously distributed component
extending less than $1.75$\rstar\ and/or an active region at the pole. 
The rotational modulation in the lightcurve and spectra indicates the presence
of at least two compact emitting regions. Peaks in the lightcurve
indicate that the regions are unlikely to be located higher than 0.4\rstar\ above
the stellar surface. 

We have shown that while we still do not have the spectral resolution at
X-ray wavelengths to comprehensively ``map'' the coronae of cool stars,
we can trace velocity variations and use these to constrain the
positions of emitting regions in the coronae of cool stars.
Spectro-polarimetric ground-based observations of AB Dor have been carried out
simultaneously with these {\em Chandra } observations.
We will use Zeeman Doppler imaging techniques to map the surface magnetic field
of AB Dor at this epoch, we will extrapolate these surface maps to produce
detailed 3D coronal magnetic field and X-ray emission models.
The X-ray emission model will be
tested against the results presented here and will be the subject of a future paper.  

\begin{acknowledgements}

The authors acknowledge the data analysis facilities provided by the Starlink P
roject which is run by CCLRC on behalf of PPARC. This work 
has been carried out thanks to the support of a Harvard Smithsonian CfA
fellowship and an ESA Internal fellowship. GAJH would also like to thank 
R. Hoogerwerf for useful discussions.

\end{acknowledgements}

\end{document}